\documentclass[aps,preprint]{revtex4}%
\usepackage{amsfonts}
\usepackage{amsmath}
\usepackage{amssymb}
\usepackage{graphicx}%
\begin{document}
\title{Sommerfeld's quantum condition of action and the spectra of Schwarzschild
black hole }
\author{L. Liu}
\email{liuliao1928@yahoo.com.cn}
\affiliation{Department of physics, Beijing Normal University}
\author{S. Y. Pei}
\email{psy@bnu.edu.cn}
\affiliation{Department of physics, Beijing Normal University}
\preprint{7 pages, no figures, submitted to Classical and Quantum Gravity}
\preprint{Report-No: \ BNU-03-66}

\begin{abstract}
If the situation of quantum gravity nowadays is nearly the same as that of the
quantum mechanics in it's early time of Bohr and Sommerfeld, then a first step
study of the quantum gravity from Sommerfeld's quantum condition of action
might be helpful. In this short paper the spectra of Schwarzschild black
hole(SBH) in quasi-classical approach of quantum mechanics is given. We find
the quantum of area is $\frac{8\pi}{3}l_{p}^{2}$, the quantum of entropy is
$\frac{2\pi}{3}k_{B}$ and the Hawking evaporation will cease as the black hole
reaches its ground state $m=\frac{1}{2\sqrt{3}}m_{p}.$

\end{abstract}
\maketitle

Several different approaches to quantum gravity have been proposed in the
recent years. Among them, people believe that loop quantum gravity [LQG]
\ might be a viable\ one to quantum gravity\cite{1}. However there are several
discrepancies with it\cite{2}. The first one, for example, is the appearance
of an undetermined free parameter---the so-called Immirzi--parameter-----which
plaques with the overall important coefficient " one quarter" in classical
black hole physics. The second one is that it has no reasonable low energy and
high energy approximation, for example, the low energy approximation of LQG is
not the Einstein's general relativity, which has a sounded experimental
support now. And LQG in the Planck regions agrees well with Einstein's general
relativity, though we have no reasons to believe it. So it seems, we are now
in the time as the old quantum physics was in the Bohr-Sommerfeld's time.
Therefore a similar first step treatment of quantum gravity by using
Sommerfeld's quantum condition of action might be a viable road to future
quantum gravity.\bigskip

\section{Action variable,\ Action and Hamilton}

\ As is known in classical mechanics, one can define a quantity called action
variable for a single periodic system\cite{3},%

\begin{equation}
I_{v}=\oint pdq \label{1}%
\end{equation}
which is really the area of phase space occupied by the system in one periodic
motion. We remind there is an important relation between the action I and the
action variable I$_{v\text{ }}$of the single periodic motion, i.e\cite{3}%
\cite{4}.%

\begin{equation}
I=I_{v}-\oint Hdt\label{2}%
\end{equation}
where H is the Hamiltonian of the system. In the so-called (3+1) decomposition
of any Einstein gravitational system, we have however two different kinds of
Hamiltons, i.e. the one $H(h_{ij},\pi_{ij})$ without \ Gibbons-Hawkijng
surface term that satisfies the famous Hamilton constraint
\begin{equation}
H(h_{ij},\pi_{ij})=0\label{3}%
\end{equation}
the other

\bigskip%
\begin{equation}
H^{^{\prime}}=H+\alpha\label{4}%
\end{equation}
where

$\bigskip$%
\begin{equation}
\alpha=\underset{r\rightarrow\infty}{\lim}\int_{s}(\frac{\partial h_{ij}%
}{\partial x^{j}}-\frac{\partial h_{jj}}{\partial x^{i}})r^{i}\label{5}%
\end{equation}
is just the ADM mass $m$ for asymptotic flat SBH. We remind $h_{ij}$ is the
induced metric of the 3-D space-like hypersuface and $\pi_{ij}$ is the
canonical conjugated momentum of $h_{ij}$. The important thing in (4) we would
like to emphasize is that the Hamilton $H^{^{\prime}}$ of any classical
Einstein gravitational system with asymptotic flateness is always the ADM mass
$m$ of the system. So an important result is for any Einstein single periodic
gravitational system with boundary, we have%

\begin{equation}
I=I_{v}-\oint H^{^{\prime}}dt=\oint pdq-m\oint dt=\oint pdq-m\mathbf{T}%
\end{equation}
where $\mathbf{T}$ is the perieod of the single periodic motion, m is the ADM
mass of the system.

\section{\bigskip Sommerfeld's quantum condition of action and the spectra of
SBH}

Early in the year of 1916, A Sommerfeld successfully put forth the so-called
Sommerfeld's quantum condition of action\cite{3}%

\begin{equation}
\oint pdq=nh
\end{equation}
in order to solve the spectra problem in atom physics. Landau \& Lifshitz
derived the Sommerfield's quantum condition for the one dimensional periodic
motion of a particle by using the quasi-classical method and get\cite{6}%

\begin{equation}
\oint pdq=2\pi\hbar(n+\frac{1}{2})
\end{equation}
that is, for any cyclic motion the action variable $I_{v}\ $of the system is quantized.

Now we try to apply this principle to SBH. As is known, the Euclidean Kruskal
section of SBH is a cyclic or single periodic system, whose metric reads

$\bigskip$%
\begin{equation}
ds^{2}=\frac{32}{r}m^{3}\exp\{-\frac{r}{2m}\}(dT^{2}+dR^{2})+r^{2}d\Omega
_{2}^{2},\ \ (r>2m)
\end{equation}
where%

\begin{equation}
iT=(\frac{r}{2m}-1)^{\frac{1}{2}}\exp\{\frac{r}{4m}\}\sin(\frac{\tau}{4m}),
\end{equation}%
\begin{equation}
R=(\frac{r}{2m}-1)^{\frac{1}{2}}\exp\{\frac{r}{4m}\}\cos(\frac{\tau}{4m})
\end{equation}
Obviously, both $T$ and $R$ in the parametric form (10) and (11) are periodic
function of $\tau$ with period $8\pi m$.\ Now we would like to emphasize, that
this peculiar property of EuclideanKruskal metric is very important to reveal
the thermodynamics of SHB. We shall see now in the following , that this is
also of key importance for the recognition of it's quantum property.

From classical general relativity, we know that the area $A$ of the event
horizon of SBH and it's action $I$ are%

\begin{equation}
A=16\pi m^{2}\ \bigskip(G=C=1)
\end{equation}
and%
\begin{equation}
I=4\pi m^{2}=\oint pdq-8\pi m^{2}\bigskip(G=C=\hbar=1,\mathbf{T=}8\pi m\text{
in eq(4)})
\end{equation}
respectively.

Here we note, for the vacuum Euclidean Kruskal section the volume integral of
it's action is zero, so the contribution to action comes only from the
Gibbons-Hawking's surface term which is $4\pi m^{2}$. From (12) and (13) the
variation $\triangle A$ of $A$ and $\triangle I$ of $I$ have the relation%

\begin{equation}
\triangle A=4\triangle I\text{ }(G=C=\hbar=1)
\end{equation}
Now after noting for SHB $H^{^{\prime}}=m,$ $I=I_{v\text{ }}-8\pi m^{2},$ we
can apply Sommerfeld's quantum condition (8) to (12) and (13) and get the
spectrum of action $I$, the spectrum of area $A$ of event horizon and the
spectrum of entropy $S$ of SBH as follows%

\begin{equation}
\bigskip I=4\pi m^{2}=2\pi\hbar(n+1/2)-8\pi m^{2}%
\end{equation}
or%
\begin{equation}
m^{2}=\frac{1}{6}(n+\frac{1}{2})m_{p}^{2}%
\end{equation}%
\begin{equation}
A=16\pi m^{2}=\frac{8\pi}{3}(n+1/2)l_{p}^{2}%
\end{equation}
and%

\begin{equation}
S=\frac{1}{4}A
\end{equation}
where%

\begin{equation}
I=S=\frac{1}{4}A
\end{equation}
.The minimum variation or quantum of the area of the event horizon $\delta A$,
quantum of the entropy $\delta S$ and quantum of the mass $\delta m$ of SBH are%

\begin{equation}
\delta A=\frac{8\pi}{3}l_{p}^{2}%
\end{equation}%
\begin{equation}
\delta S=\frac{2\pi}{3}k_{B}%
\end{equation}
and%
\begin{equation}
\delta m=\frac{1}{12}\frac{m_{p}^{2}}{m(T)}%
\end{equation}
respectively. Where $l_{p}=(G\hbar C^{-3})^{\frac{1}{2}}$ is the Planck
length, $l_{p}^{2}=G\hbar C^{-3}$ is the Planck area, $m_{p}=(G^{-1}\hbar
C)^{\frac{1}{2}}$ is the Planck mass and m(\textit{T}) is the mass of SBH at
temperature \textit{T}.

It seems our area quantum (20) only refers to the event horizon. We have no
reason to infer that this is a general result to all area of any surface.
Let's point out, our result of the quantum area (20) is different from the
recent value of $4\sqrt{3}l_{p}^{2}$ by Dreyer and Motl\cite{2}.\ The total
mass-energy loss of a SBH via Hawking evaporation in temperature $T$ is%

\begin{equation}
\int\rho(\nu,T)d\nu=\int\frac{8\pi h\nu^{3}}{\exp(\frac{h\nu}{kT})-1}%
d\nu\equiv E(T)
\end{equation}
From (20 ) and (23) we obtain the quantum of mass loss(QML) of a SBH via
Hawking evaporation at temperature T{}%

\begin{equation}
\triangle m=E(T)=\frac{1}{12}\frac{m_{p}^{2}}{m(T)}%
\end{equation}
If the black hole mass $m(T)$ has a lowest limit $\frac{1}{2\sqrt{3}}m_{p}$
(see Eq. (25) ), then the QML $\triangle M=E(T)$ has an upper limit $\triangle
m=\frac{1}{2\sqrt{3}}m_{p}$.

\section{Discussion:}

From (15 ) and (17) we see n=0 should correspond to the ground state of SBH.
It is easy to show the ground state mass m$_{G}$ of SBH is%

\begin{equation}
m_{G}=\frac{1}{2\sqrt{3}}m_{p}%
\end{equation}
It seems, there is no way to decrease the mass of a SBH under it's ground
state mass $m_{G}.$ Therefore even Hawking evaporation will cease as a SBH
approaches it's ground state! That means, no loss of quantum coherence and no
violation of unitarity in the last stage of evolution of a black hole! In a
word, the long unsolved information puzzle doesn't exist at all\cite{7}! This
is a very interesting result from our quasi-classical quantum mechanic approach.

\begin{acknowledgments}
We thank prof J. Y. Zhu for let us know the papers in ref[2] and wish to thank
Prof. Z. Zhao and Dr H. Y. Wang for helpful discussions.
\end{acknowledgments}

\end{document}